\documentclass[11pt,twoside]{article}
\usepackage{cozumel2005}
\usepackage{epsf}
\usepackage{psfig}
\begin{document}

\title{Near Infrared Survey of Populous Clusters in the LMC: Preliminary
Results} 

\author{A.J. Grocholski$^{1,2}$, A. Sarajedini$^1$, K.A.G. Olsen$^3$, and
G.P. Tiede$^4$}

\affil{$^1$Department of Astronomy, University of Florida, 211 Bryant
Space Science Center, Gainesville, Florida 32611, aaron@astro.ufl.edu,
ata@astro.ufl.edu}
\affil{$^2$Visiting Astronomer, Cerro Tololo Inter-American
Observatory}
\affil{$^3$Cerro Tolo Inter-American Observatory, National Optical
Astronomy Observatory, Casilla 603, La Serena, Chile,
kolsen@ctio.noao.edu} 
\affil{$^4$Department of Physics and Astronomy, Bowling
Green State University, Bowling Green, OH 43403, gptiede@bgnet.bgsu.edu}

\begin{abstract} %%% Abstract to run on from here.

We report preliminary results from our near-infrared JHK survey of star
clusters in the LMC. The primary goals of the survey are to study the
three-dimensional structure and distance of the LMC.  In 2003 and 2004 we
used the Infrared Side Port Imager (ISPI) on the CTIO 4m to obtain
near infrared photometry for a sample of populous LMC clusters.  We
utilize the K-band luminosity of core helium burning red clump (RC) stars
to obtain individual cluster distances and present a preliminary
assessment of the structure and geometry of the LMC based on a subset of
our data.

\end{abstract}

%%% MAIN BODY OF TEXT GOES HERE. CONSULT  manual_cozumel2005.tex 
%%% SECTIONS 2.3-2.6 FOR HELP WITH EQUATIONS, FIGURES,
%%% AND TABLES.

\section{Introduction} 

The Large Magellanic Cloud (LMC) is an attractive astronomical target for
a variety of reasons.  Owing to its relative proximity, stellar
populations in the LMC can be easily resolved. These populations exhibit
an array of star formation processes and episodes in a dynamic environment
making the LMC well suited for studying the formation and evolution of a
satellite galaxy.  Traditionally, the LMC has been thought of as an
approximately planar galaxy that, in spite of its proximity, can be
assumed to lie at a single distance from us.  In contrast, Caldwell \&
Coulson (1986) have shown that the LMC disk is tilted with respect to the
plane of the sky.  More recent work has not only confirmed that the LMC is
tilted, but it also indicates that the LMC disk is considerably thicker
than previously assumed.

van der Marel \& Cioni (2001) determined, through the use of red giant
branch and asymptotic giant branch stars as relative distance indicators,
that the LMC is tilted $34\fdg7 \pm 6\fdg2$ with respect to the line of
sight ($0\deg$ is face-on) such that the Northeast portion of the LMC is
closer to us than the Southwest.  Olsen \& Salyk (2002) confirmed this
result, finding $i = 35\fdg8 \pm 2\fdg4$ by utilizing core He burning red
clump (RC) stars as their relative distance indicator.  Additionally, by
studying carbon star kinematics in the LMC disk, van der Marel et
al.~(2002) have determined that $v/\sigma = 2.9 \pm 0.9$, implying that
the LMC disk is thicker than the Milky Way thick disk ($v/\sigma \approx
3.9$).  Lastly, the Magellanic Stream (e.g.~Putman et al.~2003), flaring
(Alves \& Nelson, 2000) and elongation of the LMC disk (van der Marel \&
Cioni 2001) and the possibility that the LMC disk is warped (Olsen \&
Salyk 2002, Nikolaev et al.~2004) all indicate that the LMC has not 
escaped unharmed from its
tidal interactions with the Milky Way and Small Magellanic Cloud.

The distance to the LMC has been a topic of considerable discussion in
recent years and a variety of methods have been employed to calculate this
distance; e.g.~variable stars (Cepheids, RR Lyraes, Miras),
color-magnitude diagram (CMD) features (main sequence turn off, tip of the
red giant branch, RC stars), and SN 1987a.  
There has been, until recently, little agreement between the different
methods and sometimes even amongst distances calculated using a single
method.  This lead to a ``long" and ``short" distance scale for the LMC
with a ``short" distance modulus, $(m-M)_0$, of $\sim$18.2-18.3 mag and a
``long" distance of $\sim$18.5-18.7 mag.  Clementini et al.~(2003)  
demonstrate this distance problem (top panel, their Fig.~8), and find that
the long and short distance scale can be reconciled, at least to within
the errors, with improved photometry and/or reddening estimates (bottom
panel, their Fig. 8) for some of the previous works.

A primary reason for interest in the LMC distance is its use as the
extragalactic distance scale zeropoint.  The Hubble Space Telescope Key
Project to determine $H_0$ (Freedman et al.~2001) utilized a sample of LMC
Cepheid variables to define the fiducial period-luminosity relation.  
Cepheid distances were then used to calibrate secondary standard candles
which lie further along the extragalactic distance ladder.  Thus, the
accuracy of their determination of $H_0$ ($72 \pm 8$ km s$^{-1}$
Mpc$^{-1}$) hinges on the accuracy of their zeropoint, $(m-M)_{0,LMC} =
18.5 \pm 0.10$.  It turns out the error in their calculation is dominated
by the uncertainty in $(m-M)_{0,LMC}$; it takes up $6.5\%$ of their $9\%$
error budget (Mould et al. 2000).

In this paper we will present preliminary results from our near-infrared
survey of populous clusters in the LMC.  Section 2 presents our
observations of LMC cluster and field stars.  In the next two sections we
discuss our application of the $K$-band luminosity of the RC as a standard
candle for calculating absolute cluster distances (\S 3) and for
determining relative distances to the LMC fields (\S 4).  Finally, in \S 5
we talk about our future work on this project.

\section{Data}
\subsection{Observations} 

We have obtained near infrared images for a sample of intermediate age LMC
clusters over the course of two, three night observing runs (20-22 January
2003 and 06-08 February 2004) at the CTIO 4m.  Our observations were made
with the Infrared Side Port Imager (ISPI) which utilizes a 2048 $\times$
2048 pixel HAWAII 2 HgCdTe array.  In the f/8 configuration, ISPI yields
an 11$\arcmin$ $\times$ 11$\arcmin$ field of view and a plate scale of
$\sim$ 0$\farcs$33 pixel$^{-1}$.  For our observations we used a
nine-point dither pattern, centered on each cluster, and total integration
times as follows:  J = 540s, H = 846s, K$'$ = 846s.  Average seeing for
all six nights was $\sim$ 1.2$\arcsec$.  Table 1 lists two of our 18 
program clusters along with right ascention and declination (J2000) and
passbands in which the clusters were observed.

\begin{table}[!ht]
\caption{LMC Cluster Sample}
\smallskip
\begin{center}
{\small
\begin{tabular}{lccccccccccc}
\tableline
\noalign{\smallskip}
Cluster & Alternate name & RA (J2000) & Dec (J2000) & Filters \\
\noalign{\smallskip}
\tableline
\noalign{\smallskip}
NGC 1651 & & 04$^h$37$^m$32$\fs$67 & -70$\deg$35$\arcmin$07$\farcs$7 &
JHK$'$ \\
Hodge 4 & SL 556 & 05 32 25.00 & -64 44 12.0 & JHK$'$ \\
\noalign{\smallskip}
\tableline
\end{tabular}
}
\end{center}
\end{table}

\subsection{Reduction}

All data was processed using standard data reduction steps, which we will
now summarize.  Images were dark subtracted, sky subtracted and then flat
fielded using on-off dome flats.  Due to the combination of ISPI's wide
field of view and the relatively large steps in our dither pattern ($\geq$
30$\arcmin$), these images suffer from geometric distortions, caused
mostly by the curvature of the focal plane.  To correct for this, we apply
a high order distortion correction to each image using the IRAF task
GEOTRAN.  The corrected images are then aligned, shifted and averaged to
create a final science frame for each cluster and filter.

Science frames were photometered using a combination of DAOPHOT and
ALLSTAR (Stetson 1987) as follows.  A rough PSF was constructed using the
brightest $\sim$ 200 stars in each image.  The rough PSF was then used to
remove neighbors from around the PSF stars, allowing the creation of a
more robust PSF from the cleaned image.  ALLSTAR was utilized to fit this
improved PSF to all stars in the science frame.  In an effort to detect
and photometer faint stars and/or companions, we used a single iteration
of subtracting all stars detected and fit in the first ALLSTAR pass, then
searching for previously undetected stars in our fields.  All new
detections were run through ALLSTAR using the same PSF as in the first
ALLSTAR pass and these stars were added to the photometry list.  At this
point, aperture corrections, calculated for each frame, were applied to
the PSF photometry.  Finally, aperture corrected photometry lists from
each filter were combined with the requirement that a star be detected
in all filters for it to be kept in the final combined list of
instrumental magnitudes.  Zero points and color tranformations appropriate
for our data were calculated by comparing our instrumental magnitudes with
photometry from the 2MASS All-Sky Data
Release\footnote{http://www.ipac.caltech.edu/2mass/releases/allsky} for
each field in our program.

\section{Preliminary Distances}

Figure 1 presents ($K$, $J-K$) CMDs for NGC 1651 and Hodge 4, where all
stars within $\sim 1\arcmin$ of the cluster centers are shown.  Both CMDs
show a prominent RC at $K \sim 16.9$ and a well populated RGB extending up
to $K \sim 12.5$.

\begin{figure}[!ht]
\plotone{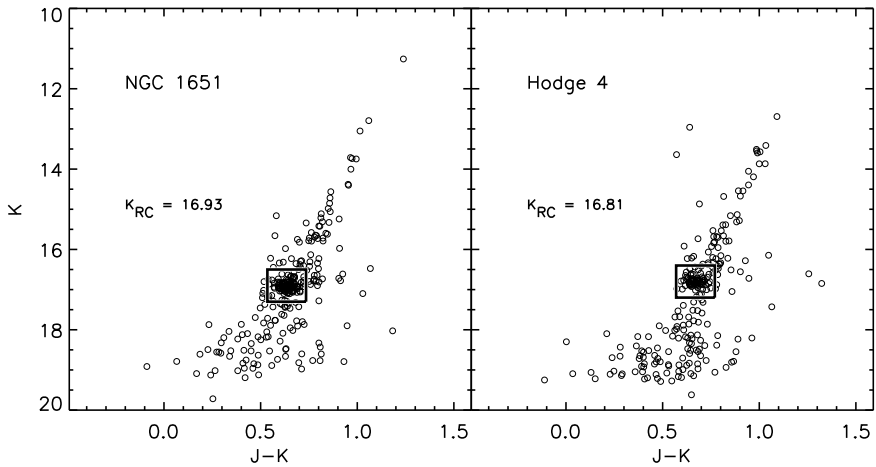}
\caption{Near infrared CMDs for NGC 1651 and Hodge 4.  Both plots show a
well populated helium burning RC, as denoted by the box, along with an RGB
extending $\sim 4.5$ mag brightward of the RC.  Median values of $K_{RC}$ 
are given.}
\end{figure}

We follow the method of Grocholski \& Sarajedini (2002) in using a box
that extends 0.8 mag in $K$ and 0.2 mag in $J-K$ (shown in Fig. 1),
centered by eye, to select the RC stars.  $K_{RC}$ is calculated by taking
the median value of all stars within this box.  For NGC 1651, $K_{RC} = 
16.93 \pm 0.02$ and $K_{RC} = 16.81 \pm 0.02$ for Hodge 4.

With regards to the reddening of each cluster, we utilize the dust maps of
Burstein \& Heiles (1982) and Schlegel, Finkbeiner, \& Davis (1998).  
Since the values determined from both dust maps, for each cluster, are in
good agreement, we adopt the average value from the two maps as our
cluster reddenings.  For NGC 1651 we find $E(B-V) = 0.12 \pm 0.02$ and for
Hodge 4, $E(B-V) = 0.05 \pm 0.01$.  Using the relations from Cardelli,
Clayton, \& Mathis (1989), $A_V = 3.1E(B-V)$ and $A_K = 0.11A_V$, these
reddenings translate to $A_K = 0.041 \pm 0.003$ and $A_K = 0.017 \pm
0.007$.

Previous authors have shown that the absolute RC magnitude varies as a
function of age and metallicity for visible and near-infrared bands, with
this variation seen in both theoretical (e.g.~Salaris \& Girardi 2002)  
and observational data (e.g.~Sarajedini 1999, Cole 1998).  An in-depth
comparison of the absolute $K$-band RC magnitude with age and metallicity
for a sample of simple stellar populations was performed by Grocholski \&
Sarajedini (2002).  These authors advocate using an interpolation over
either their observational data or the theoretical models of Girardi \&
Salaris (2001) to create an $M_K(RC)$ ``plane" which, given a cluster's
age and metallicity, can be used to predict $M_K(RC)$ for that cluster.

In many cases, however, age and metallicity values for LMC clusters are
not readily available or not reliable.  For example, Olszewski et
al.~(1991) have presented the only large scale determination of
metallicities for LMC clusters, based on the Ca II triplet.  However, many
of their cluster [Fe/H] values, including NGC 1651 and Hodge 4, are
based on observations of a single star. Sarajedini et al.~(2002)
calculated [Fe/H] values for both of these clusters using the slope of
the RGB.  While their value for Hodge 4 is consistent with that of
Olszewski et al.~(1991), their value for NGC 1651 is 0.3 dex more metal
rich.  As such, in the current work we choose not to apply the full
calibration of $M_K(RC)$ as discussed above, but rather we adopt the value
of $M_K(RC) = -1.61 \pm 0.04$ given in Grocholski \& Sarajedini (2002).  
We note that Sarajedini et al.~(2002), utilizing the full RC treatment
from Grocholski \& Sarajedini (2002), find $M_K(RC) = -1.56 \pm 0.12$ for
NGC 1651 and $M_K(RC) = -1.64 \pm 0.17$.  This implies that error in the
value we have chosen to use for $M_K(RC)$ is likely larger than that
quoted.

Using the values listed above for $K(RC)$, $M_K(RC)$, and $A_K$ for each
custer, we find for NGC 1651, $(m-M)_0 = 18.50 \pm 0.06$ and $(m-M)_0 =
18.40 \pm 0.05$ for Hodge 4.  The errors quoted are the random errors
added in quadrature.  These numbers are consistent with the LMC distance,
$(m-M)_0 = 18.50 \pm 0.10$ used in the HST Key Project to determine an
accurate value of $H_0$ (see Freedman et al.~2001 for more information).  
Additionally, these distances agree with the LMC geometry determined by
van der Marel \& Cioni (2001) and Olsen \& Salyk (2002) in that Hodge 4
should be closer to us than NGC 1651, based on the tilt of the LMC's disk
and the location of the clusters in the LMC.  Lastly, Sarajedini et
al.~(2002)  find, for NGC 1651 and Hodge 4, $(m-M)_0 = 18.55 \pm 0.12$ and
$18.52 \pm 0.17$.  These distances are in agreement, within the errors, 
with our results.

\section{Field Stars}

In Figure 2 we show ($K$, $J-K$) CMDs for the field stars surrounding our
clusters and, as with the clusters, the RC and RGB for the two fields are
easily visible.  The major difference between the cluster and field CMDs
is the wider field RGB (spread in color) and larger field RC (spread in
both color and luminosity) caused by the intrinsic distribution in age and
metallicity of the field population.  Although this situation is a bit
more complicated for the RC (see Salaris \& Girardi 2002; Grocholski \&
Sarajedini 2002), in general, older and/or more metal rich populations
have redder RGBs and redder and fainter RCs than young and/or metal poor
populations.  Due to the mixed population in the LMC field, dealing with
the field RC luminosity as a standard candle becomes a much more
formidable task than dealing with a simple stellar population RC.

\begin{figure}[!ht]
\plotone{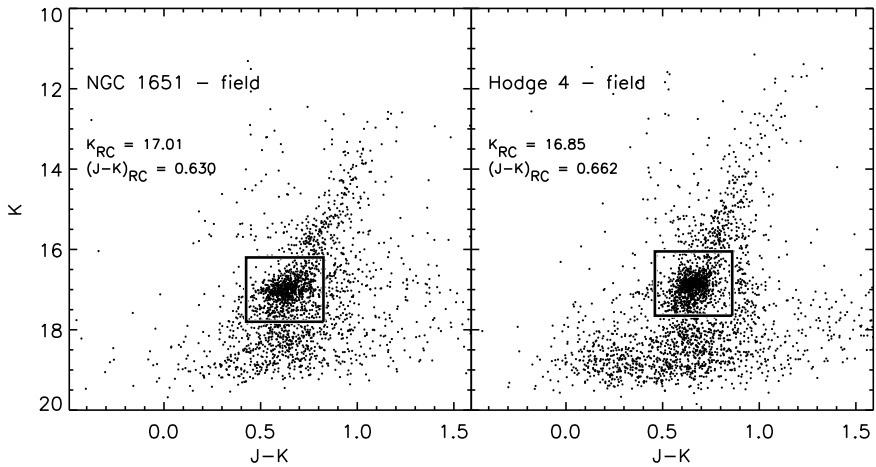}
\caption{Near infrared CMDs for field stars around NGC 1651 and Hodge 4.
Both plots show a RGB with a larger color spread and a RC with a larger
spread in both color and magnitude than the cluster CMDs.  This difference
in the CMDs is due to the intrinsic distribution in age and metallicity in
the LMC field.}
\end{figure}

However, if we can make the assumption that each observed field within the
LMC has a similar mix of stars in terms of age and metallicity, then the
field RC can be easily applied as a relative distance indicator since
$M_K(RC)$ should be the same for all fields.  This assumption is
reasonable given that the bulk of the LMC RC stars are $\sim 4$ Gyr old
(Girardi \& Salaris 2001) and differential rotation should destroy any
record of the initial local age and metallicity distribution on timescales
much shorter than this (Olsen and Salyk 2002).  Additionally, if this
assumption holds true, the variation in RC colors amongst our fields is
indicative of the relative reddenings of these fields.

Similar to \S 3, we determine the magnitude and color values of the field
RC by taking the median value of all stars within a predefined box,
centered on the RC.  Since the field RC is more extended than the cluster
RCs, we have doubled the size of our box to 1.6 mag in $K$ and 0.4 mag in
$(J-K)$, as shown in Fig. 2.  For our NGC 1651 and Hodge 4 fields,
respectively, we find $K_{RC} = 17.01 \pm 0.01$, $(J-K)_{RC} = 0.630 \pm
0.002$ and $K_{RC} = 16.85 \pm 0.01$, $(J-K)_{RC} = 0.662 \pm 0.002$.  
Applying the reddening corrections we derived from the dust maps (section
3), we find $K_0(RC) = 16.97 \pm 0.01$ and $K_0(RC) = 16.83 \pm 0.01$ for
the NGC 1651 and Hodge 4 fields.  This result implies that the field
around NGC 1651 is $0.14$ mag farther from us than the Hodge 4 field,
consistent with the results from \S 3.  However, assuming the difference
in color is solely due to a change in the reddening between these two
fields, the Hodge 4 field suffers from $E(B-V) = 0.061$ more extinction
than the NGC 1651 field, in the opposite sense of what is predicted by
both dust maps discussed in \S 3.  This could indicate that the field
populations are in fact not composed of the same stellar mixture (the
Hodge 4 field may be older and/or more metal rich), which would render
this relative distance method invalid for our data.  It is also possible
that there exists a problem in our photometric calibration, leading to the
discrepency between our results and the reddening maps of Schlegel et
al.~(1998) and Burstein \& Heiles (1982).  We are currently exploring the 
cause of this disagreement.

\section{Future Work}

As mentioned previously, the two clusters presented here are only a sample
of our entire data set.  We plan to use these data to address two
problems.  First, we will utilize $K(RC)$ values for each cluster, along
with the full method of determining $M_K(RC)$, to calculate the individual
cluster distances.  (We note that we are in the process of measuring
homogeneous ages and metallicities for each of our clusters, which will be
used in properly calculating $M_K(RC)$)  This will allow us to determine
an accurate distance for the LMC.  Additionally, with the areal coverage
of our clusters, we will be in a position to compare the distribution of
our program clusters with the LMC geometry derived by van der Marel and
Cioni (2001).  Second, the plethora of field stars surrounding each of our
clusters provides an opportunity to study the field populations in the
LMC.  If the assumption that RC stars are evenly mixed throughout the LMC
is confirmed, then we will be able to determine the relative distance to
each of these fields, thereby studying the three dimensional distribution
of the LMC disk.  This information will provide insight into the reality
of the ``warp" found in the southwest portion of the LMC disk (Olsen and
Salyk 2002).  We will also be in a position to compare the relative
distributions of our LMC clusters with their surrounding fields and
explore whether or not the LMC clusters occupy the same plane as the LMC
disk.

%%\subsection{The first subsection}
%%\subsubsection{The first sub-subsection}
%%\section*{The First Un-Numbered Section}
%%\subsection*{The first un-numbered subsection}

\acknowledgements             
This research is supported by NSF CAREER grant AST-0094048 to A. Sarajedini.

%%% THE BIBLIOGRAPHY
%%%
%%% CONSULT SECTION 3 OF   manual_cozumel2005.tex    FOR HOW TO USE NATBIB.
%%% AUTHORS ARE ENCOURAGED TO USE EITHER THE "THEBIBLIOGRAPY" ENVIRONMENT
%%% or THE BIBTEX ENVIRONMENT. 


\begin{thebibliography}{}
\bibitem[]{} Alves, D., \& Nelson, C. 2000, \apj, 542, 789
\bibitem[]{} Burstein, D., \&, Heiles, C.  1982, \aj, 87, 1165
\bibitem[]{} Caldwell, J. A. R., \& Coulson, I. M.  1986, \mnras, 218, 223
\bibitem[]{} Cardelli, J. A., Clayton, G. C., \& Mathis, J. S.  1989, 
\apj, 345, 245
\bibitem[]{} Clementini, G., Gratton, R., Bragaglia, A., Carretta, E.,
DiFabrizio, L., \& Maio, M.  2003, \aj, 125, 1309
\bibitem[]{} Cole, A. A.  1998, \apj, 500, L137
\bibitem[]{} Freedman, W.L., et al.  2001, \apj, 553, 47
\bibitem[]{} Girardi, L., \& Salaris, M.  2001, \mnras, 323, 109
\bibitem[]{} Grocholski, A. J., \& Sarajedini, A. 2002, \aj, 123, 1603
\bibitem[]{} Mould, J. R., et al.  2000, \apj, 529, 786
\bibitem[]{} Nikolaev, S., Drake, A. J., Keller, S. C., Dalal, N., Griest, 
K., Welch, D. L., \& Kanbur, S. M.  2004, \apj, 601, 260
\bibitem[]{} Olsen, K.A.G., \& Salyk, C.  2002, \aj, 124, 2045
\bibitem[]{} Olszewski, E. W., Schommer, R. A., Suntzeff, N. B., \& 
Harris, H. C.  1991, \aj, 101, 515
\bibitem[]{} Putman, M. E., Staveley-Smith, L., Freeman, K. C., Gibson, B.
K., \& Barnes, D. G.  2003, \apj, 586, 170
\bibitem[]{} Sarajedini, A.  1999, \aj, 118, 2321
\bibitem[]{} Sarajedini, A., Grocholski, A. J., Levine, J. L. \& Lada, E. 
A.  2002, \aj, 124, 2625
\bibitem[]{} Salaris, M., \& Girardi, L.  2002, \mnras, 337, 332
\bibitem[]{} Schlegel, D. J., Finkbeiner, D. P., \& Davis, M.  1998, \apj, 
500, 525
\bibitem[]{} Stetson, P. B. 1987, \pasp, 99, 191
\bibitem[]{} van der Marel, R. P., \& Cioni, M-R. L.  2001, \aj, 122, 1807
\bibitem[]{} van der Marel, R. P., Alves, D. R., Hardy, E., \& Suntzeff, 
N. B.  2002, \aj, 124, 2639

%\bibitem[]{}
%\bibitem[]{}
%\bibitem[]{}
%\bibitem[]{}
%\bibitem[]{}
%\bibitem[]{}
%\bibitem[]{}
%\bibitem[]{}
\end{thebibliography}
\end{document}